# Automatic Liver Segmentation from CT Images Using Deep Learning Algorithms: A Comparative Study

K. E. Şengün, Y. T. Çetin, M.S Güzel, S. Can and  E. Bostancı[*]

*Abstract*— **Medical imaging has been employed to support medical diagnosis and treatment.  It may also provide crucial information to surgeons to facilitate optimal surgical preplanning and perioperative management. Essentially, semi-automatic organ and tumor segmentation has been studied by many researchers. Recently, with the development of Deep Learning (DL) algorithms, automatic organ segmentation has been gathered lots of attention from the researchers. This paper addresses to propose the most efficient DL architectures for Liver segmentation by adapting and comparing state-of-the-art DL frameworks, studied in different disciplines. These frameworks are implemented and adapted into a Commercial software, "LiverVision".  It is aimed to reveal the most effective and accurate DL architecture for fully automatic liver segmentation. Equal conditions were provided to all architectures in the experiments so as to measure the effectiveness of algorithms accuracy, and Dice coefficient metrics were also employed to support comparative analysis. Experimental results prove that "U-Net" and "SegNet" have been superior in line with the experiments conducted considering the concepts of time, cost, and effectiveness. Considering both architectures, "SegNet" was observed to be more successful in eliminating false-positive values. Besides, it was seen that the accuracy metric used to measure effectiveness in image segmentation alone was not enough. Results reveal that DL algorithms are able to automate organ segmentation from DICOM images with high accuracy. This contribution is critical for surgical preplanning and motivates author to apply this approach to the different organs and field of medicine.**

## I. INTRODUCTION

### A. Motivation

Estimation of organs from Computed Tomography (CT) plays an important role during the choice of treatment strategies for most organ diseases [1]. Despite the recent development in Computer Vision and Machine Learning fields, organ segmentation is still a challenge due to the difficulty in discriminating the edges of organs coupled with high variability of both intensity patterns and anatomical appearances. All these difficulties become more prominent, especially in pathological organs. The need for CT scan analysis for pre-diagnosis and treatment of internal organs is increasing due to the expectation in developed countries. Automated organ estimation of an internal CT scan can help radiologists analyze scans faster and estimate organ images with fewer errors. One of the biggest problems with the use of computed tomography is the selection of the area to be screened larger than the targeted area in most abdominal examinations. This is a major obstacle for doctors to make accurate and rapid progress in the clinical diagnosis and treatment of patients. Segmentation of the organ to be examined from the whole computed tomography image makes a great contribution to the effective examination of these large-scale abdominal images. Several studies are carried out on this subject with deep learning practices that have gained importance in recent years [3,4]. Providing organ and tissue segmentation in CT images has long been one of the topics of deep learning technology in the literature [5]. By considering these current studies, it is aimed to propose a comprehensive study to support the studies in the field of automatic segmentation by using Deep Learning algorithms. It should be stated that results provide a preliminary preparation to segment the liver more quickly and accurately. The new findings that will emerge can thus facilitate the treatment and intervention of patients faster and more effectively for doctors. The algorithm's effectiveness and prediction process have a more dominant role in practical clinical life in automated segmentation processes. Therefore, the structure to be created should be presented as fast and accurate as possible, and the rate of misleading the end-user should be minimized. Examination of liver segmentation with different amounts of data, architectures, and implementation methods may be a critical step for progress on this understanding. What encourages authors to do this study is the desire to inform researchers about the subject and the comparison made on the algorithms can be guiding and supportive for future studies. The proposed algorithms will be integrated into "LiverVision" software, a commercial medical tool, which has emerged as an indispensable tool for medical image processing, enabling to produce 3D anatomic models of the liver already [11]. An example output from the software is illustrated in Figure 1, allowing automated 3D construction and visualization of livers and blood vessels obtained from CT images.

K.E. Şengün and M.S Güzel is with Robotic Lab of Computer Eng. Dept. of Ankara University, TR (e-mails:eylulsengun@gmail.com, mguzel@ankara.edu.tr.
Y. T. Cetin is Design Engineer at Turkish Aerospace, 06980, Ankara, TR (e-mail: tugbacetin@yahoo.com), S. Can is a Senior Software Engineer at Akgun R&D Center , (e-mail: canserhat91@gmail.com), E Bostanci  is with SAAT Laboratory of Computer Eng. Dept. of A.U (e-mail: ebostanci@ankara.edu.tr)

*B. Related Works*

Image segmentation can be evaluated on two basic concepts. The first is semantic segmentation and the other is instance-based segmentation. Semantic segmentation has been studied recently and it basically aims to recognize and classify images by considering pixel levels [2]. Deep learning has provided a significant impact on the field of Semantic segmentation as it is expected [7]. Several comprehensive Image Segmentation framework has been developed in the literature in recent times. U-Net, created by O. Ronneberger, has been very popular among these frameworks, and is a good example of Convolution Neural Networks (CNNs) based semantic segmentation solution to the field of medical image segmentation.-The earliest version of Unet framework represented drawbacks.

developed for labelling cells from light microscopic images and is able to produce quite accurate segmentation results [8] A recent study claims that U-Nets provides better effective recovery of high frequency edges in "sparse-view" CTs [9]. Another study employed U-Net framework to extract retinal vessels, which is critical for the diagnosis of fundus diseases [10]. However, the drawbacks brought by U-Net can be quite insufficient in preventing the complexity caused by close

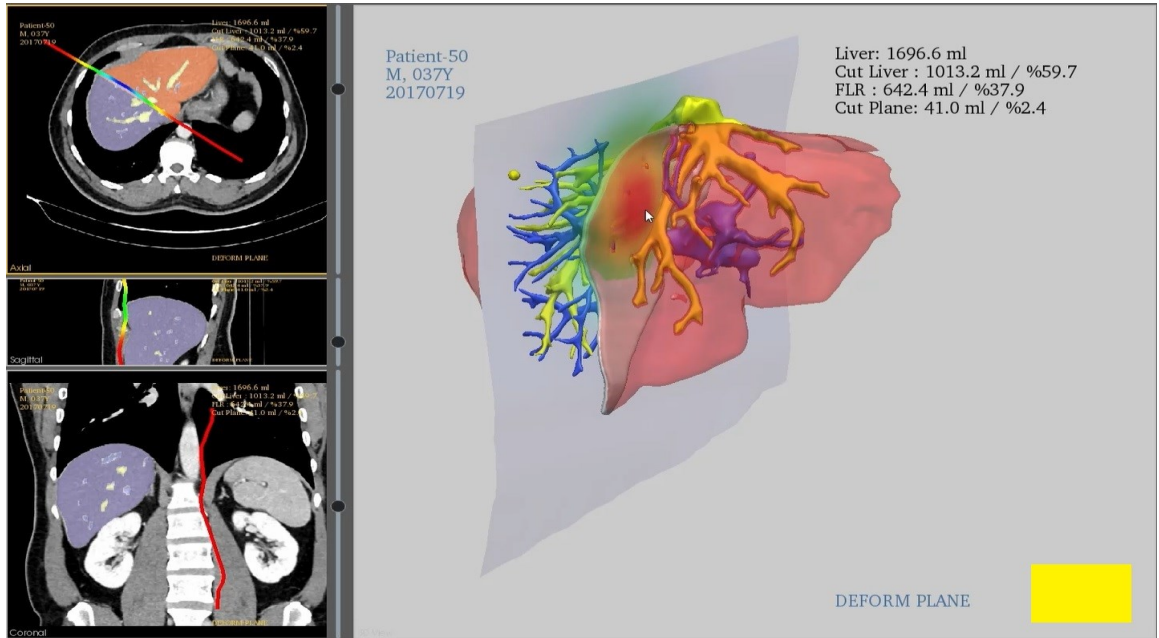

Fig. 1. An example Output from LiverVision Software, Dr. Guzel's team was contributed to the development

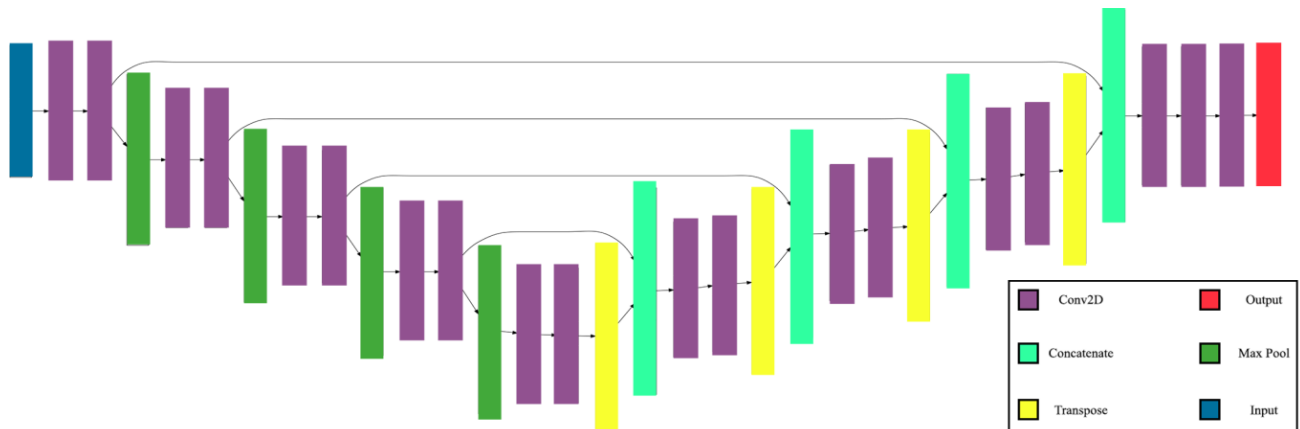

Fig. 2. The classical U-net architecture (32x32 pixels in the lowest resolution, and 512x512 in the highest resolution).

For this reason, improvements are carried out based on the implementation of classical U-Net framework. Accordingly, new and more effective algorithms are designed and implemented. One of the leading of these studies that will be evaluated in this study is the "Dense U-Net" framework, can be seen in Figure 3. This framework involves an algorithm that works more effectively in extracting more adequate features and eliminating false-positives in different tissue structures such as the color intensity of the long-distanced organs [12,13].

Besides, "ResNet" has been employed, which is a deeper layered architecture, allowing the residual learning system to facilitate the training of deeper networks that are difficult to train [14]. This network has two pooling operations, convolution layers with "3×3" and "1 × 1" filters, followed by a fully connected layer, and a Softmax classifier. The final one is the "SegNet" Architecture which is able to perform pixel-wise segmentation more effectively [15]. There are no fully connected layers and hence it has only convolutional layer. According to which, a decoder upsamples its input using the transferred pool indices from its encoder to produce a sparse feature map(s). It then performs convolution with a trainable filter bank to densify the feature map. The final decoder output feature maps are fed to a soft-max classifier for pixel-wise classification, as illustrated in Figure 4.

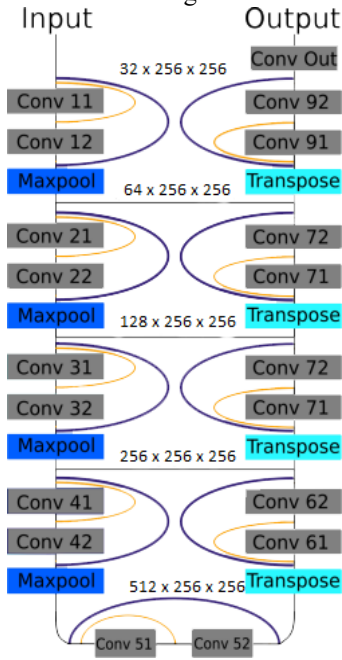

Fig. 3. Dense U- Net Architecture resolution.

As aforementioned, the first contribution of this study to the field is to develop a DL based automatic liver segmentation tool, which has been integrated into a commercial medical imaging software. In order to achieve this contribution, state-of-the-art DL frameworks will be adapted for the problem. Accordingly, a comparative comparison between those algorithms will be carried. This assessment will be the second contribution of this study. The following section will mainly introduce the corresponding methodology, covering aforementioned steps respectively.

## II. METHODOLOGY

This study basically follows three main stages to perform a comparative comparison among aforementioned architectures to develop a DL based organ segmentation architecture, integrated into the "LiverVision" software.

The first is to perform pre-processing operators to enhance the CT images. The second stage is the adaptation of algorithms to the problem, considering equal conditions. The third and final stage is to examine the results by using quality metrics and predict sample images. Each stage will be detailed as follows:

### A. Pre-Processing Steps of CT Images

The CT images of anonymous patients are employed during the experiment due to the patient privacy. The dataset consists of 20 patients' torso CT images and liver masks of these images. The masking process was carried out by experts on the subject by the party providing the tomography images. These masks allow the liver to be separated from other organs in the segmentation and the training processes. CT images and masks are in 512x512 resolutions. However, depending on the structure of the architectures used and the feature of the hardware, it is improbable to use all the images or to remain stable at the resolution. Elimination of distortions, glare and noises in the pictures and bringing a meaningful data are among the important factors that will increase the success rate in segmentation processes on the picture. One of the biggest problems in medical images (with CT or similar images) is the noise obtained from low resolution images and lack of detail. Gaussian Based Smoothing [16] and Anisotropic Diffusion Filter methods [17] are applied to handle these problems within the procedures of the pre-processing stage of this study.

Gaussian based smoothing process is a fast and effective isotropic (evenly distributed) picture smoothing method with successful results in removing the noises in the picture. The method employs the Gaussian distribution. Gaussian distribution equation is seen in (1). The biggest disadvantage of this distribution is that the large dispersion is isotropic, in which case the losses that are contained in the corner (edge) occur [18].

$$G(x,y) = \frac{1}{2\pi\sigma^2} e^{-\left(\frac{x^2+y^2}{2\sigma^2}\right)}$$
(1)

Here "$x$" refers the distance from the origin in the horizontal axis, "$y$" refers the distance from the origin within the vertical axis, and "$\sigma$" denotes the standard deviation of the Gaussian based distribution. On the other hand, Anisotropic Diffusion Filter is an anisotropic softening method that produces very successful results in removing the noises in the images. Unlike the Gaussian smoothing method, the anisotropic softening method preserves the corner points during the process. Thus, the noises on the image are eliminated and the image is softened, while the sharp points in the image are preserved [17, 19]. The corresponding equations is given below:

$$I_t = div(f(\|\nabla I\|)\nabla I)$$
(2)

Here, ‖∇I‖ illustrates the scale of gradient to be applied to the image, whereas, the function f in the formula (3) is expressed as a diffusion function and is used to determine the corner points in the images.

$$f(\|\nabla I\|) = \frac{1}{1+\left(\frac{|\nabla I|}{s}\right)^2}$$
(3)

With the corner stop function expressed in Equation 3, it is decided whether the points on the picture are vertices. Here "s" is a fixed parameter and is taken as the corner sensitivity value.

### B. The Frameworks used for Automatic Segmentation

#### 1) U-Net

The main purpose of U-Net's emergence is its use in biomedical image segmentation. It takes its name from the shape of the structure of the algorithm. It consists of an encoder network followed by a decoder network. The biggest difference of this algorithm from the classification algorithms is that it reflects the area on the pixels when the discriminatory features are learned [12]. It won the "Grand Challenge for Computer-Automated Detection of Caries in Bitewing Radiography at ISBI 2015", and it also won the "Cell Tracking Challenge at ISBI 2015" on the two most challenging transmitted light microscopy categories [20]

#### 2) Dense U-Net

Dense U-Net has a deep-layered structure. Each layer takes the input maps of the previous layers as input and transfers them to the next layers. Thus, a higher segmentation result can be obtained by using fewer data [21]. Despite U-Net is considered as a modern CNN based medical segmentation model, it is also claimed that it is not strong enough to capture advanced details. The Dense U-Net enjoys the benefits of both the U-net and Dense-net. It employs dense concatenations to extend the depth of the contracting path [22]. The model is also illustrated in Figure 3.

#### 3) ResNet

The ResNet variation adapted for the experiments is a residual network consisting of 18 layers. Apart from segmentation, it is also used to increase the performance of many applications such as computer vision or facial recognition. It is considered that the use of multi-layered and deep structures will provide a straight proportional increase in the accuracy and success of the training. However, contrary to common belief, as the structure of the architecture deepens, there is a fluctuation between the saturation in training success and the problem of overfitting.

More layers can lead to more training errors. This confirms that not all architectures can be optimized correctly and easily. Also, layered structures can lengthen the training period in image processing and segmentation. Therefore, instead of using the more layered variations of ResNet in our experiments, we have provided an equal environment that is more convenient for comparison.

#### 4) SegNet

SegNet is an effective architecture when used for pixel-wise segmentation. It implements a process called decoding. This decoding process has a denser structure compared to U-Net. This gives SegNet many advantages: it improves boundary delineation, reduces the number of parameters needed for training. Also, its structure can be easily adapted to any encoder-decoder architecture.

SegNet is able to perform "pixel-wise" segmentation as U-Net can perform, but there are some obvious differences between them. SegNet relies on feature maps transferred from encoder to decoder instead of using pooling indices. Despite this may increase the size of the model and therefore need more memory, it may also allow to achieve more effective results [15].

### C. Quality Measure

During all conducted experiments, two performance parameters are employed, namely, *Accuracy* and *Sorensen's Dice coefficient*. Accuracy is concerned with the correct classification of pixels within the image, the equation is given in (4). In this way, it provides the ratio of parameters not suitable for training on the created masks and their CT images. However, accuracy is not solely adequate to measure success in image segmentations. On the other hand, Sorensen's Dice coefficient formula, indicated in Equation 5, is considered as a reliable assessment metric in image segmentation. This coefficient, used as a quality metric, excludes false-positives and positives that the algorithm does not compare, which makes it possible to examine the results of the wrong tissue selection that may occur in the liver segmentation.

$$Accuracy = \frac{TP + TN}{TP + TN + FP + FN} \quad (4)$$

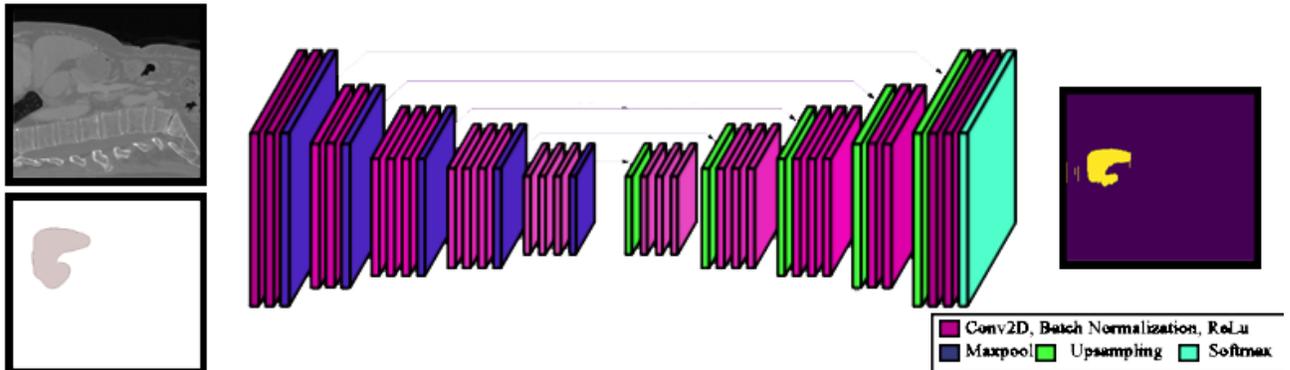

Fig. 4. An illustration of the SegNet architecture.

Here, TP refers True Positive, TN is for True Negative, FP and FT indicate False Positive, False Negative respectively. Once, the original form is adapted to the Boolean data Sorensen's Dice coefficient is illustrated as follows (5):

$$DSC = \frac{2TP}{2TP + FP + FN} \quad (5)$$

TABLE 1
VALID U-NET TEST CASES

| Cases (Train and Validation) | Number of CT Slices / Patients | Number of Epochs | Batch Size | Validation Split Rate |
|---|---|---|---|---|
| Case 1 | 3739 / 13 | 20 | 10 | 0.2 |
| Case 2 | 4385 / 15 | 6 | 16 | 0.2 |
| Case 3 | 4385 / 15 | 10 | 16 | 0.2 |
| Case 4 | 5750 / 20 | 5 | 16 | 0.4 |

TABLE 2
VALID DENSE U-NET AND RESNET TEST CASES

| Cases (Train and Validation) | Number of CT Slices / Patients | Number of Epochs | Batch Size | Validation Split Rate |
|---|---|---|---|---|
| Dense U-Net | 3739 / 13 | 10 | 10 | 0.2 |
| ResNet | 3739 / 13 | 15 | 10 | 0.2 |

TABLE 3
VALID SEGNET TEST CASES

| Cases (Train and Validation) | Number of CT Slices / Patients | Number of Epochs | Batch Size | Validation Split Rate |
|---|---|---|---|---|
| Case 1 | 2998 / 10 | 10 | 16 | 0.2 |
| Case 2 | 4385 / 15 | 10 | 16 | 0.2 |
| Case 3 | 4385 / 15 | 15 | 16 | 0.2 |
| Case 4 | 4385 / 15 | 20 | 16 | 0.2 |

## III. EXPERIMENTS

These section details the experimental process, system configuration, the results based on statistical comparison and the influence of model parameters on model validation.

### A. Pre-training

For the pre-training process, all CT images and masks have been converted to the Neuro Imaging Informatics Technology Initiative (NIfTI) format, which is a common format for processing. The main purpose is to obtain the liver images of the patients easily in a grouped manner. Hence, additional information and titles in the metadata in CT images are ignored the fact that NIfTI is in a format that simplifies this structure has provided us convenience.

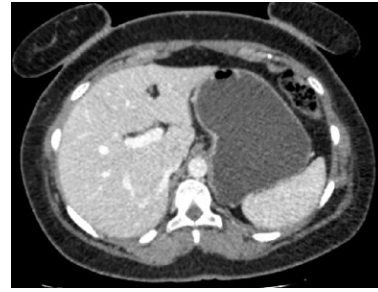

Fig. 5. An example slice of a patient's abdominal CT image. The dense area in the upper left corner is the liver.

Considering the training cost of the deep architectures, the resolution of the CT images in the data set is reduced to "256x256". The images were kept in pixel sequence format for easy processing and intervention during their implementation. In this format, the pixels were wrapped in an 8-bit unsigned integer data type. This data type allows the image to be transferred to the background without losing its original structure. Data centering and normalization processes were performed respectively, using the mean and standard deviations of the data.

### B. Implementations

All the architectures are compared under the same experimental conditions in order to provide a fair comparison. The hardware has a Linux environment consisting of 8 cores and 32 disks with 56 GiB RAM. To reduce the noisy data and make the training more balanced, Adam optimization function are implemented for all architectures Adam's value is fixed to "$(1e) - 3$" in all experiments. The implementations are generated using Keras API functions along with TensorFlow. The parameters we manipulated in the comparison are the amount of data, the number of epochs, the batch size, and the division rate of the dataset for validation.

Image processing and segmentation may result in higher computing costs than their counterparts based on other deep learning studies. When deep-layered architectures are also considered, classical architectural approaches with early stop functions are preferred. In this way, it is aimed to minimize the possibility of encountering the overfitting situation during and after the training. Besides some further adjustments are carried out by using regularization functions to prevent overfitting due to the random state of the models at their initial weight.

During the experimental process two critical problems have been encountered. The first one, overfitting, caused by memorization in the segmentation processes of the models and the second one, undersampling, which the non-learning situation is caused by the small training set according to the structure of the deployed architectures. Also, this article discussed superficially about the situations of overfitting, undersampling, and out of memory situations arising from training examples working in an iteration. Many experiments have been conducted to obtain a fair comparison, however, in the results section of the paper, the best results obtained from some of the experiments have been presented. The parameters and data quantities in these test cases are illustrated in Tables 1, 2, and 3. A gentle cross validation is employed that for each case training data is randomly selected. An *epoch* is reading of the entire dataset once. *Batch size* tells you the number of train samples obtained in one iteration. *The validation split rate* is the rate of separation of 2D images sliced in a vertical direction obtained from 3D CT images of the total patient data as train/validation sets. It is observed the final status of the models that emerged as a result of these test cases using the quality metrics and NIfTI files consisting of predicted masks. The tests presented in aforementioned tables involve cases that have achieved remarkable success in the Dice coefficient and prediction results. During the verification process, other patient files that are excluded from the training and the validation processes, are employed to verify practical predictions.

### C. Results

Based on the progress made during the epoch period, we have revealed the quality metric results of the training processes mentioned earlier. With the models generated, the predictions made with the mentioned hardware took approximately one minute per patient. The slice in Fig. 5 belongs to the patient CT image used in the prediction phase. This slice is the 124$^{th}$ image captured out of a total of 125 2-D images vertically along the sagittal. One of the most crucial points in the comparison is the relationship between validation data and training results. The closer the training results to the validation results, the better the practical success of the model generated. Besides, situations such as the validation results exceeding the train results, or the results are not in the 0-1 range indicate that an accurate model training process has not taken place.

#### 1) Experimental results obtained from U-Net

As can be seen in Table 4, we obtained the best results in Case 1. Since the accuracy values of the cases are very close to each other, the Dice coefficient metric is more suitable for comparison in this situation. The outcome that can be formed here is that when the prediction is performed, the model produced in "Case 1" is anticipated to yield a more accurate result with less false-positive results, illustrated in Figure 7.

TABLE 4
U-NET EXPERIMENT RESULTS:
FINAL QUALITY METRIC VALUES OF U-NET TEST CASES

| Cases (Train and Validation) | Final Train Accuracy | Final Validation Accuracy | Final Train Dice Coefficient | Final Validation Dice Coefficient |
|---|---|---|---|---|
| Case 1 | 0.9978 | 0.9936 | 0.9812 | 0.9260 |
| Case 2 | 0.9935 | 0.9907 | 0.8799 | 0.7707 |
| Case 3 | 0.9970 | 0.9903 | 0.9479 | 0.8271 |
| Case 4 | 0.9929 | 0.9665 | 0.8719 | 0.4385 |

Figure 6 illustrates the variation of model parameters for U-Net architecture.

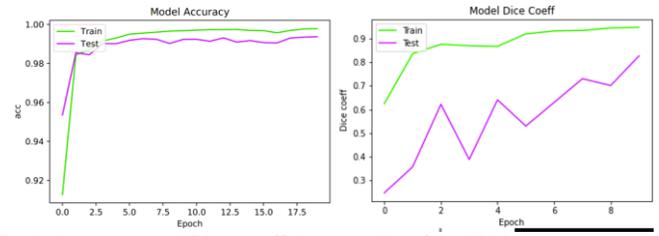

Fig. 6. Accuracy and Dice coefficient graphs of the U-Net experiment

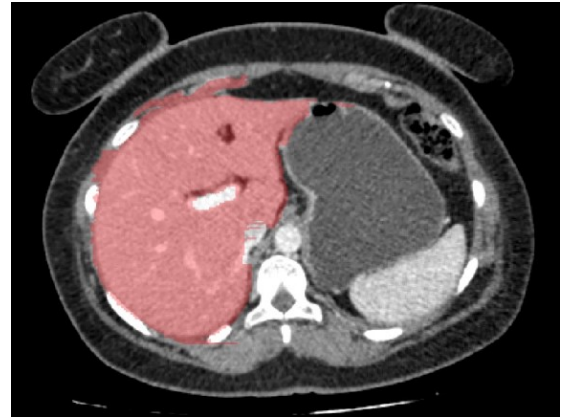

Fig. 7. Sample prediction images represented with Case 1 of U-Net experiments on slice, shown in Fig. 5 (Liver part is shown with red)

It can be stated that Case 1, shown in Fig. 7, yields the closest output to the liver image, presented in Fig. 5 However, some markings have emerged in addition to the segmentation process in the area where the liver is located. This shows that the biggest drawback that stands out in the U-Net architecture compared to other algorithms is that it tends to mark areas that does not belong to the liver. These areas may confuse the end-user during the practical use of the segmentation process. On the other hand, once such markings are neglected, the user may face serious problems.

#### 2) Experimental results obtained from Dense U-Net vs ResNet

ResNet and Dense U-Net tend to memorize instead of improving the results when changes are made to the amount of data and epoch due to their deep layered structure. Training and validation length is about two times longer than other algorithms. (According to the configuration of the hardware used during experiments, an epoch length of the architectures remains under 20 minutes for U-Net and SegNet, whereas this time can exceed 45 minutes for Dense U-Net and ResNet.) Table 5 illustrates the final quality metrics for Dense U-Net and ResNet, illustrated in Table 2.

TABLE 5
DENSE U-NET AND RESNET EXPERIMENT RESULTS: FINAL QUALITY METRIC VALUES OF DENSE U-NET AND RESNET

| Cases (Train and Validation) | Final Train Accuracy | Final Validation Accuracy | Final Train Dice Coefficient | Final Validation Dice Coefficient |
|---|---|---|---|---|
| Dense U-Net | 0.9953 | 0.9914 | 0.9659 | 0.8890 |
| ResNet | 0.9975 | 0.9939 | 0.9787 | 0.9241 |

Results reveal that ResNet performed better than U-Net when dealing with false-positive values. However, some positive areas of liver images have also been ignored due to their strict structure, which requires more extensive examination. This led to the selection of a much smaller area than the prediction operation should be carried out. On the other hand, Dense U-Net acts as an intersection set of these features of ResNet and U-Net. The acceptance of the organs away from the liver as positive areas decreased, and its correct detection of more positive areas in the liver increased. However, for close organs, the margin of error in false-positive increased compared to these algorithms (see Fig. 8). ResNet achieved a more successful result in terms of both the values seen in Table 5 and the reasons stated above.

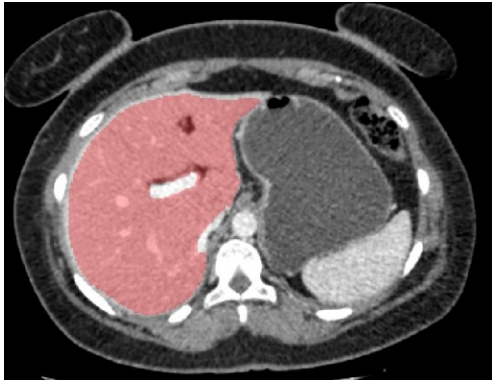
(a)

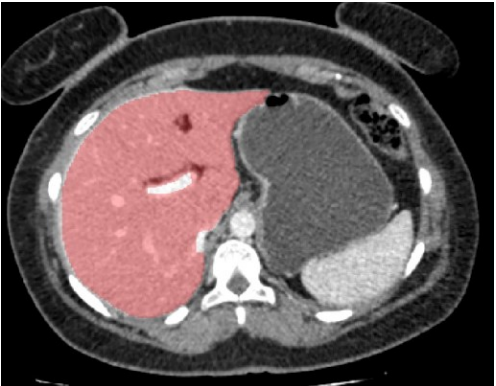
(b)

Fig. 8 Visualization of Sample prediction results from the slice of Dense U-Net (a) and ResNet (b) (Liver parts are shown with red)

### 3) Experimental results obtained from SegNet

The resulting SegNet model has revealed a smaller model size than ResNet and Dense U-Net, which carried out experiments under the same environmental conditions. That is because, SegNet does not have a fully convolutional structure despite its decoding structure. Since U-Net has a less compact layer structure, the size of the model created with this architecture is smaller than SegNet. Also, since SegNet's training period is shorter than the others, it was easy to modify model parameters during training process. Fig 9. shows the fluctuations that Case 3 experienced, in line with the epochs. It should be noted that, SegNet is more flexible in responding to data growth without overfitting compared to other algorithms.

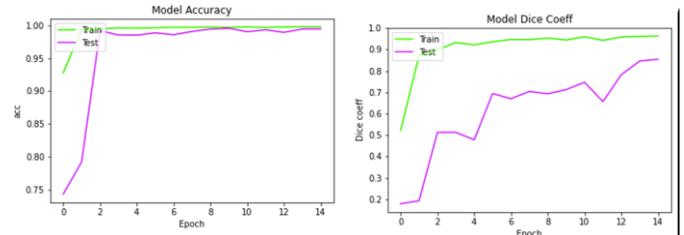

Fig. 9. Accuracy and Dice coefficient graphs of the SegNet experiment the lower and upper limit values of the x and y axes of the graphs depend on the number of epochs during training

TABLE 6
SEGNET EXPERIMENT RESULTS: QUALITY METRIC VALUES OF SEGNET

| Cases (Train and Validation) | Final Train Accuracy | Final Validation Accuracy | Final Train Dice Coefficient | Final Validation Dice Coefficient |
|---|---|---|---|---|
| Case 1 | 0.9974 | 0.9812 | 0.9429 | 0.8007 |
| Case 2 | 0.9972 | 0.9908 | 0.9517 | 0.7544 |
| Case 3 | 0.9978 | 0.9944 | 0.9636 | 0.8548 |
| Case 4 | 0.9984 | 0.9905 | 0.9735 | 0.8061 |

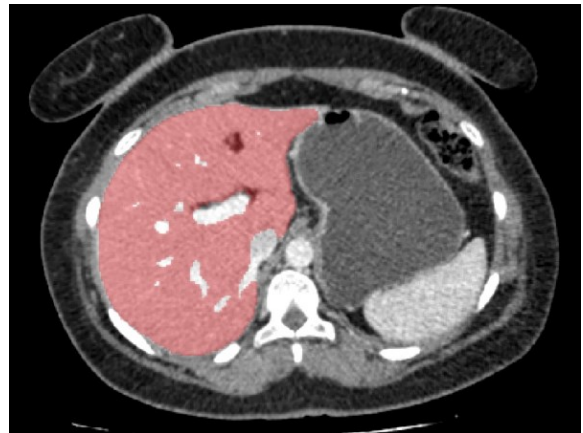

Fig. 10. Sample prediction images of SegNet experiment cases on slice in Fig. 5 (Liver parts are shown with red)

According to Table 6, the most successful model belongs to Case 3. Although the U-Net Dice score is more successful than this algorithm, SegNet is practically more successful in eliminating false-positives and detecting positives more comprehensively. The reason why U-Net gives a higher Dice coefficient than SegNet, it is more prone to approaching false-positive areas as positive. 3-D view of a ground truth liver mask used as reference in this study illustrated in Figure 11.

4) Statistical comparison and evaluation

Previous results reveal the superiority of models using SegNet and U-Net architectures. Hence, a paired t-test is applied to the results so as to make a comparison between these models in a statistical manner. According to the results, it has been observed that SegNet yields better results than U-Net in practice, and so to confirm this, we first performed predictions on the CT images of 12 patients by employing these architectures. Accordingly, to measure the similarity between them, the "Hausdorff" distance [26] between is employed by considering ground truth liver mask, shown in Figure 11. The hypothesis tried to be proved here is that whether the distance measurement of SegNet is equal to or greater than U-Net. It should be stated that the lesser the Hausdorff distance, the closer the two images are to each other. Consequently, Table 7 reveals that SegNet is closer to ground truth image.

TABLE 7
T-TEST RESULTS

| Group | U-Net | SegNet |
|---|---|---|
| Mean | 5.582775 | 5.522137 |
| SD | 1.614765 | 1.567370 |
| SEM | 0.316681 | 0.307386 |
| N | 26 | 26 |

According to the t-test results, there was a significant difference in the scores for U-Net (M=5.582775, SD=1.614765) and SegNet (M=5.522137, SD=1.567370) by considering conditions; t=2.079617, p=0.0239, α=0.05. Since p-value<α, the initial hypothesis is rejected. Also, the observed standardized effect size is medium (0.41). That indicates that the magnitude of the difference between the average and "μ0" is medium. Therefore, predictions made using the SegNet architecture produces a better result compared to the success of the U-Net architecture and achieved a significant increase.

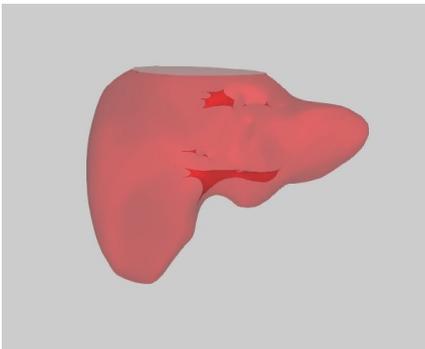

Fig. 11. 3-D view of Ground Truth Image Mask involving Polygons

5) Evaluation and influence of model parameters

After model implementations and regularization processes, the biggest success in model training processes is directly related with the estimation and analysis of the parameters. During the training process, various conjectures have emerged with the results obtained through observation. For instance, increasing the number of epochs excessively to obtain a better dice coefficient may cause the model to experience overfitting.

As it is expected, increasing the batch size shortens the model training time. Although it may seem like a positive process, some of the ignoring made to create this situation in the background affects the performance of the model quite seriously. This means that the batch size needs to be adjusted very carefully to optimize the processing time while doing model training. Increasing validation split also means less training data. Although the variety of data will decrease, the accuracy printed at the validation may not reflect the reality for the image segmentation, even if the accuracy is high at the training phase. Therefore, the relationship between image amount and validation split should be determined correctly. According to the observations, in most cases, a value between "0.2" and "0.4" works fine.

The excessive number of images may cause overfitting. However, increasing the number of images at a correct rate has a positive effect on the Dice coefficient. The reduction in the accuracy score is due to the increase in image diversity. Thus, increasing the amount of image in the right amount and variety has a positive effect on models created, on image processing and segmentation. The model using SegNet framework is adapted into LiverVision software and a final output regarding to the patient, given in Figure 5, is illustrated in Figure 12.

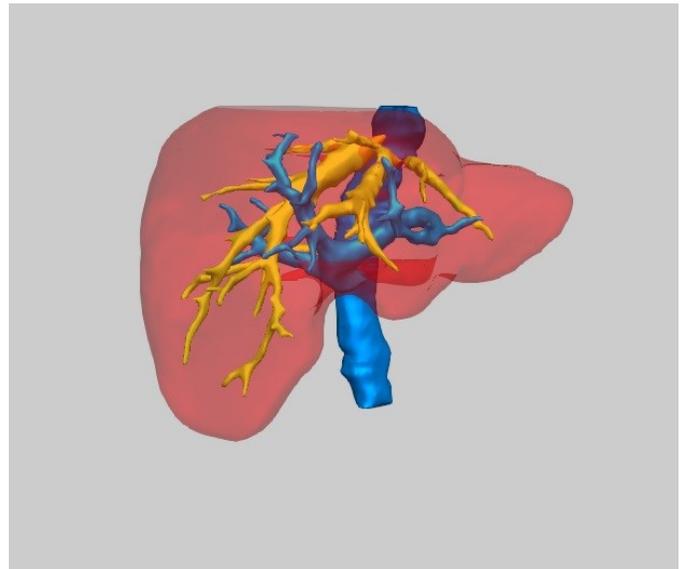

Fig. 12. An example output from LiverVision Software based on the proposed automatic liver segmentation model

IV. CONCLUSION

This study performs a comparative analysis on automatic liver segmentation problem based on Deep Learning models.

Accordingly, the success of different algorithms used in many different areas, including organ and tissue estimation on CT images on liver segmentation are evaluated. State-of-the art Deep learning architectures from different disciplines, involving, U-Net, Dense U-Net, ResNet, and SegNet, are adapted for automatic liver segmentation problem. The ultimate goal was to implement and adapt the most suitable architecture into a Commercial software, "LiverVision".

Through several comprehensive experiments, it is observed that the proposed model based on SegNet architecture has shown a more successful performance for automatic liver segmentation, considering the time, cost, practical prediction success, and effectiveness. Consequently, the most critical contribution of this study is to design and implement a novel automatic liver detection module. Accordingly, as aforementioned a model using SegNet architecture is implemented and integrated into the "LiverVision" software. In addition to this contribution, a general understanding of the effects of some critical model parameters, data amount, and quality metrics used in deep learning on the models for medical imaging are comparatively and critically discussed. It is believed that these results and analysis will be supportive to those who will work in the field of medical imaging in the future during their architectural implementations.

Deep learning, automatic organ segmentation, and fast prediction can cause a lot of computing costs on machines. That is the reason why the prediction algorithms that exist in different implementation methods should be improved. Such an improvement will provide quick and effective decisions in the diagnosis and treatment of the disease. Thus, deep learning can find more places in the medical world in practice.

## DATA AVAILABILITY

A public dataset, available in NifTi format, is employed and can be reached by the following link:

https://github.com/soribadiaby/Deep-Learning-liver-segmentation

## ACKNOWLEDGMENT

We would like to thank co-founder and facilitator of Medi Vision software company, Mr. Cumhur Çeken for his endless support and collaboration throughout this study.

## CONFLICT OF INTERESTS

The authors declare that there are no conflicts of interest regarding the publication of this paper.